\documentclass[letterpaper, 10pt, conference]{ieeeconf}  

\IEEEoverridecommandlockouts                              
\usepackage{graphics} 
\usepackage{epsfig} 
\usepackage{mathptmx} 
\usepackage{times} 
\usepackage{amsmath} 
\usepackage{amssymb}  
\usepackage{psfrag}
\usepackage{bm}
\newcommand{\ddt}{\frac{\textnormal{d}}{\textnormal{dt}}}

\title{\LARGE \bf
Inversion of Linear and Nonlinear Observable Systems \\ with Series-defined Output Trajectories
}

\author{Jean-Fran\c cois Stumper and Ralph Kennel\textit{, Senior Member, IEEE}
\thanks{This work was supported through the National Research Funds of Luxembourg under grant PhD-08-070.}
\thanks{J-F. Stumper and R. Kennel are with the Institute of Electrical Drive Systems and Power Electronics, Department of Electrical Engineering and Information Technology, Technische Universit\"at M\"unchen, D-80333 Munich, Germany.
{\tt\small jean-francois.stumper@tum.de}}%
}

\begin{document}

\maketitle
\thispagestyle{empty}
\pagestyle{empty}

\begin{abstract}

The problem of inverting a system in presence of a series-defined output is analyzed. Inverse models are derived that consist of a set of algebraic equations. The inversion is performed explicitly for an output trajectory functional, which is a linear combination of some basis functions with arbitrarily free coefficients. The observer canonical form is exploited, and the input-output representation is solved using a series method. It is shown that the only required system characteristic is observability, which implies that there is no need for output redefinition. An exact inverse model is found for linear systems. For general nonlinear systems, a good approximation of the inverse model valid on a finite time interval is found. 


\end{abstract}


\section{INTRODUCTION}

Generally, a dynamical system is described by a model consisting of a set of differential equations. With knowledge of the control input, the states and the output are computed by solving these equations. For many controller designs, however, a so-called \textit{inverse model} is required. Here, with knowledge of the output, the states and the input are computed, without solving differential equations. 

An intuitive application of an inverse model is a feedforward controller, where the output reference trajectory is given and the corresponding control input is generated. Other applications include trajectory tracking control, two-degree-of-freedom control, as well as trajectory optimization and inversion-based predictive control.

The methods presented in this paper are applicable if the output trajectory is defined as a series functional. In many applications, for example in robotics, it is common practice to design a continuous trajectory as power series through polynomial interpolation between desired setpoints \cite{SV89}. As this is a simple, straightforward and efficient method, and as often only the setpoints are of interest but not the path itself, this procedure has become popular also for feedforward controller trajectories. It is also common in trajectory optimization, where the path between the setpoints is also considered. Applying the Ritz-Galerkin method (or basis function approach), the series definition of the output trajectory transforms the optimal control problem to a finite-parameter optimization problem \cite{GF06} \cite{SK10}. 

Thus, inverse models and series-defined output trajectories are very common in the current state-of-the-art as well as in research in automatic control.

It is noted that the term \textit{system inversion} is slightly different as in the famous works of Devasia \cite{DCP96}, where numerical integration is involved. The aim here is to provide an inverse model in algebraic form.

Typically, the system is first inverted, then the output is defined as series functional \cite{HSR04}. Inversion is then related to differential flatness (or feedback linearization), where the output is redefined such that the system has full relative degree \cite{HSR04}. However, this redefinition, if possible, may be undesirable if the output is dependent on uncertain parameters \cite{H03}. Furthermore, tracking of the original output is made difficult  because of the dynamics between the original and the redefined output, which must be respected for reference recalculation \cite{SK09} \cite{HagSCL}. In contrast, this method mostly relies on observability, and simplifies the implementation as an arbitrary output is used.

For conciseness, only single-input single-output (SISO) systems are considered. However, the methods are directly extendable to multi-input multi-output systems.

The paper is organized as follows. In section II, the considered problem of inversion with series-defined outputs is formulated. In section III, the basic idea of exploiting observability is introduced. Section IV shows the method for linear systems, and section V treats nonlinear systems. Two examples are shown for linear systems and one for nonlinear systems.

\section{SYSTEM INVERSION WITH SERIES-DEFINED OUTPUTS}

Usually, exact system inversion is achieved through exploiting differential flatness, which can be seen as an extension to controllability \cite{HSR04}. The result is a differential parameterization of the system states and the control input
\begin{align}
\bm{x}(t) &= \bm{A} \left( y(t),\dot{y}(t),\ldots,y^{(n-1)}(t) \right)  \label{eq:diffparamx} ,  \\
u(t)          &= B \left( y(t),\dot{y}(t),\ldots,y^{(n)}(t) \right)  \label{eq:diffparamu}   .
\end{align}
Knowledge of the output trajectory $y(t)$ leads to the state and control input trajectories, without solving differential equations. In many applying controllers, subsequently, the output trajectory is defined as a series functional
\begin{align}
y(t)=\sum_{i=0}^{N} \alpha_i \Psi_i(t), \;\; t\in[t_0,t_f],  \label{eq:ritz}
\end{align}
to reduce the trajectory problem to a simpler finite-parameter problem. The arbitrary free parameters $\alpha_i$ could be determined by polynomial interpolation between setpoints, or by an optimization scheme for trajectory generation.

In this paper, the two steps of computing the inverse and defining the output trajectory are interchanged. Given a trajectory defined according to (\ref{eq:ritz}), the corresponding state trajectories $\bm{x}(t)$ and control input $u(t)$ are searched. Thus, the state and control input trajectories are determined as algebraic functions of time and the arbitrary parameters $\bm \alpha$
\begin{align}
\bm{x}(t) &= \bm{A}(\bm \alpha, t)  \label{eq:paramx}  , \\
u(t) &= B(\bm \alpha, t)  \label{eq:paramu}   .
\end{align}
This is the general form in which inverse models with series-defined outputs are applied. The difference in the procedure is of interest as the system class is not limited to differentially flat or controllable systems, and as the inversion is performed for an arbitrary output.

\section{PARAMETERIZATION BASED ON OBSERVABILITY}

Definition: Observability

\textit{A system is observable if for any $T > 0$ it is possible to determine the state of the system $\bm{x}(T)$ through measurements of $y(t)$ and $u(t)$ on the interval $[0,T]$.}\\

Thus, exploiting the property of observability for system inversion seems a good choice. The next two subsections present system-theoretic results that, in the following, lead to a solution of the inversion problem.

\subsection{Parameterization of the system states $x(t)$}

The observer canonical form for a nonlinear SISO system \cite{GB81} is given as 
\begin{align}
 \left\{ \begin{array}{l}
    \dot{x}_1 \;\;\;\; = \; x_2 \;\;\;\;+ g_1(x_1) u    \\
    \dot{x}_2 \;\;\;\; = \; x_3 \;\;\;\;+ g_2(x_1,x_2) u   \\
    \vdots    \\
    \dot{x}_{n-1}      = \; x_n \;\;\;\; +g_{n-1}(x_1,\ldots,x_{n-1}) u \\
    \dot{x}_n \;\;\;\; =  F(x) \;+g_n(x_1,\ldots,x_n) u \\
  \end{array}     \right.   \label{eq:obscan} \;, 
\end{align}

where the output is $y=x_1$. It is noted that, however, the observer canonical form can not be found for every nonlinear observable system. A parameterization of the system states through the output $y$ and input $u$ is found straightforward by successive elimination of the states $x_i$. The resulting parameterization is
\begin{align}
\begin{array}{l}
x_1 = y , \\
x_2 = \dot{y} -  g_1(y)u , \\
x_3 =\ddt x_2 -g_2(y,x_2) u , \\
   \;\; \vdots     \\
x_n  = \ddt x_{n-1} -g_{n-1}(x_1,\ldots,x_{n-1}) u .  \\  
\end{array}   \label{eq:obsparamx}
\end{align}
Computation of these equations is straightforward. Note that the last equation of (\ref{eq:obscan}) is unused.

A point to consider here are initial conditions. If a system is of full relative degree, the initial conditions are regarded on the output trajectory only. If not, as in the examples in this paper, some conditions involve the control input trajectory $u(t)$ too.

\subsection{Parameterization of the control input $u(t)$}

A further result on observability is that for every observable system, an input-output representation can be found \cite{vdS89}. Input-output representations are higher-order differential equations in the inputs and outputs, and are equivalent to state-space descriptions. The input-output equation takes the form
\begin{align}
q(y,\ldots,y^{(n)},u,\ldots,u^{(m)}) = 0, \label{eq:hodgl}
\end{align}
in which the output appears with its $n$-th and the input with its $m$-th derivative. To determine the input $u(t)$ for a given output trajectory, (\ref{eq:hodgl}) is solved. It is noted that this is equivalent to solving the last line of (\ref{eq:obsparamx}). 

Thus, for a full parameterization of the system variables, the solution of one differential equation is required. If the output is given as (\ref{eq:ritz}), the solution of the differential equation is possible in a straightforward manner with the widely known series methods \cite{TP63}. The solutions are presented in the next two sections.

\section{RESULTS FOR LINEAR SYSTEMS}

This section demonstrates that an exact inverse model is easy to compute for an arbitrary output of a linear observable system.

\subsection{Exact inversion with a power-series output}

The state trajectories $\bm{x}(t)$ are determined by successive elimination of (\ref{eq:obsparamx}). As the observer canonical form is linear, the functions $g_i(\cdot)$ are constants and the function $F(x)=\bm{q}^T\bm{x}$ is linear. The result is
\begin{align}
\begin{array}{l}
x_1 = y               , \\
x_2 = \dot{y} - g_1u  , \\
x_3 = \ddot{y}- g_1\dot{u} - g_2 u , \\
 \;\; \vdots  \\
x_n = y^{(n-1)} - g_1u^{(n-2)} - \ldots - g_{n-1}u  . \\
\end{array}   \label{eq:linobsparamx}
\end{align}
Furthermore, the input-output representation is a linear equation, as follows from the last equation of (\ref{eq:obscan}):
\begin{align}
\dot{x}_n = \bm{q}^T\bm{x} +g_n u  ,
\end{align}
and consequently,
\begin{align}
\bm{q}^T\bm{x} = y^{(n)} -g_1u^{(n-1)} -\ldots - g_{n-1}\dot{u} - g_n u  ,
\end{align}
which becomes a linear higher-order differential equation in $u$ and $y$ with (\ref{eq:linobsparamx}). For conciseness, it is rewritten as
\begin{align}
\sum_{i=1}^{n} k_{yi} y^{(i)} = \sum_{i=1}^{m} k_{ui} u^{(i)} . \label{eq:linhodgl}
\end{align}
The output trajectory $y(t)$ is defined as a power series
\begin{align}
y(t)=\sum_{i=0}^{N} \alpha_i t^i, \;\; t\in[t_0,t_f],
\end{align}
meaning in (\ref{eq:ritz}) is set $\Psi_i(t)=t^i$. Power series simplify the procedure, as $\ddt \Psi_i(t)=\Psi_{i-1}(t)$ and $\Psi_0=1$. In the examples, it will be shown that this choice is useful for observable and controllable systems, but that for non-controllable systems, other basis functions shall be used. The derivatives are
\begin{align}
y^{(i)}(t)= \sum_{j=0}^{N-i+1} \frac{j!}{(j-i)!} \alpha_{j+i} t^{j}, \;\; t\in[t_0,t_f].
\end{align}
The applied method is known as method of undetermined coefficients \cite{TP63}. Explicit computation of the initial conditions $u(t_0),..,u^{(m-1)}(t_0)$ is avoided. The control input is a power series with the undetermined coefficients $\bm \beta$,
\begin{align}
u(t)=\sum_{i=0}^{N} \beta_i t^i, \;\; t\in[t_0,t_f]. \label{eq:lininput}
\end{align}
Both trajectories are inserted in (\ref{eq:linhodgl}) to obtain
\begin{align}
\sum_{i=1}^{n} k_{yi} \sum_{j=0}^{N-i} \frac{j!}{(j-i)!} \alpha_{j+i} \ t^{j} = \sum_{i=1}^{m} k_{ui} \sum_{j=0}^{N-i} \frac{j!}{(j-i)!} \beta_{j+i} \ t^{j} .
\end{align}
The undetermined coefficients $\beta_i$ are then found comparing the terms in equal power of $t$. To find the inverse model, the parameters $\beta_i$ are eliminated and replaced in (\ref{eq:lininput}) by the linear functions of $\bm \alpha$.

Exactly $n$ of the coefficients $\alpha_i$ are fixed by the initial conditions $\bm{x}(t_0)$. Evaluation of (\ref{eq:linobsparamx}) leads to the respective equations. It is however noted that the information included in (\ref{eq:linobsparamx}) is also included in the input-output representation, thus the initial value problem could be transformed such that some of the coefficients $\beta_i$ are fixed.

The main result is that the differential equations are exactly solved. The inversion is equivalent to an exact inversion followed by an output trajectory approximation (\ref{eq:ritz}).


\subsection{Example of a linear controllable system: Buck converter}

The average model of the buck converter, or synchronous step-down converter, is given by the linear lifferential equations in the capacitor voltage $u_c$ and the inductor current $i_l$
\begin{align}
\left\{ \begin{array}{l}
C\ddt u_c = \frac{-1}{(R_C+R)} u_c + \frac{R}{(R_C+R)} i_l      \\
L\ddt i_l \ = \frac{-R}{(R_C+R)} u_c + \left(-R_L-\frac{RR_C}{(R_C+R)}\right) i_l + U_{IN} u   \\ \end{array} , \label{eq:linexample} \right.
\end{align}
where the control input $u$ is the duty cycle. The output is the voltage $y=U_{OUT}$
\begin{align}
U_{OUT} = \left(\frac{R}{R_C+R}\right) u_c + \left(\frac{R_CR}{R_C+R}\right) i_l .
\end{align}
It is noted that $u_c$ is the controller canonical form output, not $U_{OUT}$. For conciseness, the parameters are set to $U_{IN}=12$V, $C=200\mu$F, $L=100\mu$H, $R=2\Omega$, $R_C=85$m$\Omega$ and $R_L=280$m$\Omega$. The states are eliminated via the observer canonical form, yielding
\begin{align}
u_c &= 0.98y  -18.6\dot{y} +181950 u   ,  \label{eq:linexuc} \\
i_l &= 0.707y +219 \dot{y} -2143200u   .  \label{eq:linexil}
\end{align}
The input-output representation therefore is
\begin{align}
\ddot{y} +0.006 \dot{y} =  575 u + 9780 \dot{u}. \label{eq:linex-hodgl}
\end{align}

The output trajectory and the control input are defined as degree $3$ power series
\begin{align}
y(t) = \sum_{i=0}^{3} \alpha_i t^i , \;\;\;  u(t) = \sum_{i=0}^{3} \beta_i t^i,  \;\;\; t\in[0,T]. \label{eq:ulinansatz}
\end{align}

To determine the coefficients $\beta_i$, the linear system of equations when inserting (\ref{eq:ulinansatz}) in (\ref{eq:linex-hodgl}) is solved. Comparing the terms in equal power of $t$, the undetermined coefficients are found as
\begin{align}
 \begin{array}{l}
    \beta_0 = 0.00001043 \alpha_1 +0.003123 \alpha_2 -0.1593 \alpha_3 ,\\
    \beta_1 = 0.00002086 \alpha_2 +0.009369 \alpha_3  ,\\
    \beta_2 = 0.00003130 \alpha_3 ,\\
    \beta_3 = 0  .\\
   \end{array}   
\end{align}

Setting $t_0=0$, the initial conditions are $u_c(0)$ and $i_c(0)$. Evaluating (\ref{eq:linexuc}) and (\ref{eq:linexil}) fixes the first two parameters of the output trajectory $\alpha_0$ and $\alpha_1$ as
\begin{align}
\alpha_0 &= 0.9615 u_c(0) +0.08167 i_l(0) +0.3065 \alpha_2 -15.64 \alpha_3 , \\
\alpha_1 &= -0.003457 u_c(0) +0.004792 i_l(0) +34.04 \alpha_2 -1737 \alpha_3 .
\end{align}
The coefficients $\alpha_2$ and $\alpha_3$ are thus free and can be used in a trajectory generation scheme, for example reference trajectory interpolation. For an output trajectory approximated according (\ref{eq:ulinansatz}), the inverse model is found in terms of the arbitrary parameters of the output trajectory $\alpha_i$. The states are
\begin{align}
u_c &= (0.98\alpha_0 -16.70\alpha_1 +568.2\alpha_2 -28984\alpha_3) \notag\\ &\;\;+(0.98\alpha_1 -33.40\alpha_2 +1704\alpha_3)t \notag\\ &\;\;+(0.98\alpha_2 -50.10\alpha_3)t^2 +0.98\alpha_3t^3      , \\
i_l &=  (0.707\alpha_0 +196\alpha_1 -6693\alpha_2 +341411\alpha_3)    \notag\\ &\;\;+(0.707\alpha_1 +393.2\alpha_2 -20079\alpha_3)t       \notag\\ &\;\;+(0.707\alpha_2 +589.9\alpha_3)t^2    +0.707\alpha_3t^3        , 
\end{align}
and the control input is
\begin{align}
u &=   (1043\alpha_1 +31.23\alpha_2 -0.1593\alpha_3) +(2086\alpha_2 +93.69\alpha_3)t \notag\\ &\;\;+3130\alpha_3t^2 .
\end{align}


\subsection{Example of a linear uncontrollable system}

To demonstrate that the proposed inversion scheme is only based on the observability of a system, the method is now applied to an uncontrollable and thus also non-flat system. The constructed example is defined as
\begin{align}
 \left\{ \begin{array}{l}
    \dot{x}_1 =  x_2 +  u    \\
    \dot{x}_2 = -2 x_2       \\
  \end{array}     \right.   \label{eq:noncontrol} \;, 
\end{align}
with the output $y = x_1$. The system is in observer canonical form, and has an uncontrollable subsystem $x_2$. The input-output representation is given as
\begin{align}
\ddot{y} +2 \dot{y} =   \dot{u} + 2 u , \label{eq:noncontrol-hodgl}
\end{align}
where a priori, no pole-zero cancellation happens. If power series are applied, some terms will cancel in the input-output representation, same as in a transfer function in Laplace domain. The initial conditions would be restricted to $x_2(0)=0$. To avoid this, the basis functions must be chosen such that $\ddt \Psi_i (t) \neq \Psi_{i-1} (t)$. For instance, possible choices are exponential functions $\Psi_i=e^{-i\,t}$ or trigonometric series $\Psi_i=\sin(i\,t)$. Also, a combination of such functions with power series is interesting, as the uncontrollable subsystem and the input-output behavior can be designed independently.

The output trajectory and the control input are defined as the degree $3$ exponential series
\begin{align}
y(t) = \sum_{i=0}^{3} \alpha_i e^{-i\,t} , \;\;\;  u(t) = \sum_{i=0}^{3} \beta_i e^{-i\,t},  \;\;\; t\in[0,T].
\end{align}
Placing this definition into the input-output description, and comparing the terms in the respective exponent, three undetermined coefficients are found,
\begin{align}
\beta_0 &= 0 ,\\
\beta_1 &= -\alpha_1 ,\\
\beta_3 &= -3\alpha_3 .
\end{align}
The initial conditions define the first parameter of the output trajectory, as well as the missing parameter of the input trajectory
\begin{align}
\alpha_0 &= x_1(0) -\alpha_1 -\alpha_2 -\alpha_3 , \\
\beta_2 &= -x_2(0) -2 \alpha_2 .
\end{align}
The inverse model for the given output trajectory is thus found as
\begin{align}
x_1(t) &= \alpha_0 + \alpha_1 e^{-t} + \alpha_2 e^{-2t} + \alpha_3 e^{-3t} ,\\
x_2(t) &= -2\alpha_2e^{-2t} -\beta_2e^{-2t} = x_2(0)e^{-2t} ,\\
u(t)   &= -\alpha_1e^{-t} -(x_2(0)+2\alpha_2)e^{-2t} -3\alpha_3e^{-3t} .
\end{align}
A posteriori, the restriction $\beta_i=0$ could be removed, for instance by adding a term $\alpha_4t$ to $y(t)$. This result shows that the choice of basis functions $\Psi_i(t)$ is important. Which type of functions is advantageous is depending on the system characteristics as well as on the application.

\section{RESULTS FOR NONLINEAR SYSTEMS}

Successive elimination of the observer canonical form (\ref{eq:obscan}) yields the parameterization of the state trajectories $x(t)$ (\ref{eq:obsparamx}). 

For a full parameterization, the control input $u(t)$ is determined by solving the input-output representation (\ref{eq:hodgl}). The nonlinearities are inherited therein. Even though an analytic solution of this differential equation exists in many cases, this is typically an implicit solution in terms of elementary functions, which may be less useful than an explicit series solution \cite{FLC88}. 

To solve the differental equation, the method of undetermined coefficients is extended to nonlinear systems \cite{TP63}. An alternative would be the method of successive differentiation, which is however only applicable for systems with a well-defined relative degree. This method also works with singularities as $u^{(m)}$ is not eliminated. Singularities are common in non feedback-linearizable (or non-flat) systems.

With (\ref{eq:ritz}), equation (\ref{eq:hodgl}) becomes
\begin{align}
f(\bm \alpha, t,u,..,u^{(m)}) = 0. \label{eq:func}
\end{align}
Assuming that the input $u$ is defined as a series of same type as the output $y(t)$ in (\ref{eq:ritz})
\begin{align}
u(t)=\sum_{i=0}^{N} \beta_i \Psi_i(t), \;\; t\in[0,T],  \label{eq:ritzinput}
\end{align}
the input-output representation is further simplified to
\begin{align}
f(\bm \alpha, \bm \beta, t) = 0. \label{eq:funcabt}
\end{align}
The function $f$ is supposed to be analytic, meaning, $f$ has a Taylor series expansion in powers of $t$, valid in a neighborhood of $t_0$. 

As there are $N+1$ undetermined coefficients $\beta_i$, $N+1$ independent equations must be evaluated. In a nonlinear system, for instance comparing terms in equal power in $t$ when power series are used, as in the previous section, is generally not possible as the degree of (\ref{eq:funcabt}) may be higher than $N+1$. However, a similar result is obtained by evaluating the equations
\begin{align}
f^{(i)}(\bm \alpha,\bm \beta,t) = \frac{\textnormal{d}^i}{\textnormal{dt}^i} f(\bm \alpha, \bm \beta, t) =0 , \;\; i=0..N-1 . \label{eq:difffuncabt}
\end{align}
Some coefficients $\beta_i$ may be well defined by the initial conditions. For the remaining $R\le N+1$ coefficients, a Taylor series solution of (\ref{eq:funcabt}) around a point $t=t_0$ is searched. The remaining $R$ undetermined coefficients $\beta_i$ are obtained by solving the system of algebraic equations
\begin{align}
 \begin{array}{r}
  f(\bm \alpha,\bm \beta,t_0)       = 0 , \\
  f^{(1)}(\bm \alpha,\bm \beta,t_0) = 0 , \\
     \vdots  \;\;\;\;\;         \\
  f^{(R-1)}(\bm \alpha,\bm \beta,t_0) = 0 .\\
 \end{array}   \label{eq:nleqsys}
\end{align}
which is equivalent to solving the input-output representation by ignoring the higher-order terms. This method is comparable to Taylor series approximation of the input trajectory. As this is a system of nonlinear equations, finding a solution may be difficult. Thus it cannot be guaranteed whether the method is applicable for a given system.

The inverse model is exact at $t=t_0$, but generally has an error for $t\neq t_0$. The error is bounded in a finite interval around $t_0$ and is dependent on the order of the series $N$. Also, the choice of basis functions $\Psi_i(t)$ is relevant. Thus, only an inverse model of the type (\ref{eq:paramx}) (\ref{eq:paramu}) for 'short' time intervals around $t=t_0$ can generally be found.

An open question is the convergence of the input trajectory to the exact input, which is of interest to determine the required number of parameters $\beta_i$. The number of parameters for a finite interval may grow exponentially with the interval length.

\subsection{Example for approximate inversion: Van de Vusse reactor}

The Van de Vusse type reactor is presented in \cite{POD02} as a benchmark example, mainly for operating point changes of nonlinear systems. The state-space model is
\begin{align}
\left\{ \begin{array}{l}
\dot{x}_1 = -k_1x_1 -k_3x_1^2 +u(C_{A0}-x_1) \\
\dot{x}_2 = k_1x_1 -k_2x_2 -ux_2 \\ \end{array} , \label{eq:VanVusse} \right.
\end{align}
where $x_2$ represents the product concentration in the output stream (mol/L), and $x_1$ the reactant concentration inside the reactor (mol/L). The control input $u$ is the dilution rate in the input flow relative to the reactor volume (1/h), and $C_{A0}$ is the reactant concentration in the input system (mol/L). The output is the product concentration
\begin{align}
y &= x_2 ,
\end{align}
and the system is observable. Even though the system is differentially flat \cite{CSTRflat}, a flatness-based design requires output redefinition to a full relative degree output. As $x_2$ has relative degree one, there are internal dynamics associated to this output.

For conciseness, the parameters of the system are fixed to $k_1=50\frac{1}{h}$, $k_2=100\frac{1}{h}$, $k_3=10\frac{L}{mol\cdot h}$, $C_{A0}=10\frac{mol}{L}$ and $V=1L$. The states are eliminated as
\begin{align}
x_2 &= y                                       , \label{eq:vanvussestate2} \\
x_1 &= \frac{1}{50}\dot{y} +2y +\frac{1}{50}yu , \label{eq:vanvussestate1}
\end{align}
and the input-output-representation is
\begin{align}
50\ddot{y} +7500\dot{y} +250000y +10\dot{y}^2 +2000\dot{y}y +100000y^2 \notag\\
 +u(100\dot{y} +7500y +20\dot{y}y +2000y^2 -250000) \notag\\
 +u^2(10y^2+50y) +\dot{u}(50y) =0. \label{eq:CSTRio}
\end{align}
The input-output representation is a nonlinear function as $u$ is involved as square. The output trajectory and the control input are defined as power series
\begin{align}
y = \sum_{i=0}^{N} \alpha_i t^i ,\;\;\; u = \sum_{i=0}^{N'} \beta_i t^i, \;\;\; t\in[t_0,t_f], \label{eq:CSTRi}
\end{align}
with an output trajectory degree $N=3$ and with input trajectory degree $N'$ to be determined later. The system of equations (\ref{eq:nleqsys}) can now be established. 

Setting $t_0=0$, the initial conditions of the system are $x_1(0)$ and $x_2(0)$. Comparing to (\ref{eq:vanvussestate2}) and (\ref{eq:vanvussestate1}), the first parameter of the output trajectory and of the input trajectory are fixed:
\begin{align}
\alpha_0 &= x_2(0) ,  \\
\beta_0  &= 50\frac{x_1(0)}{\alpha_0} -\frac{\alpha_1}{\alpha_0} -100.
\end{align}

\subsubsection{Solvability issue}

To analyze the issue of solvability, it is seen that the undetermined coefficients $\beta_i$ are involved as shown in table \ref{tbl:involved} for $t_0=0$.

\begin{table}[h!]
\centering
\caption{Degree of the parameters in the respective equations\label{tbl:involved}}
\begin{tabular}{|l|c|c|c|c|c|}
\hline
  & $\beta_0$ & $\beta_1$ & $\beta_2$ & $\beta_3$ & $\beta_4$ \\
\hline
$f(\bm\alpha,\bm\beta,t_0)$       & 2 & \textbf{1} & 0 & 0 & 0 \\
$f^{(1)}(\bm\alpha,\bm\beta,t_0)$ & 2 & 1 & \textbf{1} & 0 & 0 \\
$f^{(2)}(\bm\alpha,\bm\beta,t_0)$ & 2 & 2 & 1 & \textbf{1} & 0 \\
$f^{(3)}(\bm\alpha,\bm\beta,t_0)$ & 2 & 1 & 1 & 1 & \textbf{1} \\
\hline
\end{tabular}
\end{table}

As the remaining coefficients $\beta_1..\beta_N$ are sequentially involved as first-order ($\beta_i$ in $f^{(i)}(\cdot)$), they can be determined by successive elimination of the equations (\ref{eq:nleqsys}). The resulting parameterization is available in the desired format $u(t) = B(\bm \alpha, t)$ , leading to $\bm{x}(t) = \bm{A}(\bm \alpha, t)$  with (\ref{eq:vanvussestate1}). The results are not printed due to space constraints. 

\subsubsection{Quality of the inverse model}

The parameterization does not solve the input-output representation for all times $t$, but merely for $t=t_0$. This is related to the nonlinearity of the equation, as in (49), the order of the input-output equation is seen to be $N'^2\times N^2$ because of the term $u^2y^2$. To solve the system, the higher-order terms are ignored. Thus, the found inverse model is not exact and the error or offset is proportional to the order of the Taylor polynomial $N'$ in (\ref{eq:CSTRi}). If more information is desired, $N'$ must be increased.

\begin{figure}[t]
  \centering
  \includegraphics[width=0.48\textwidth]{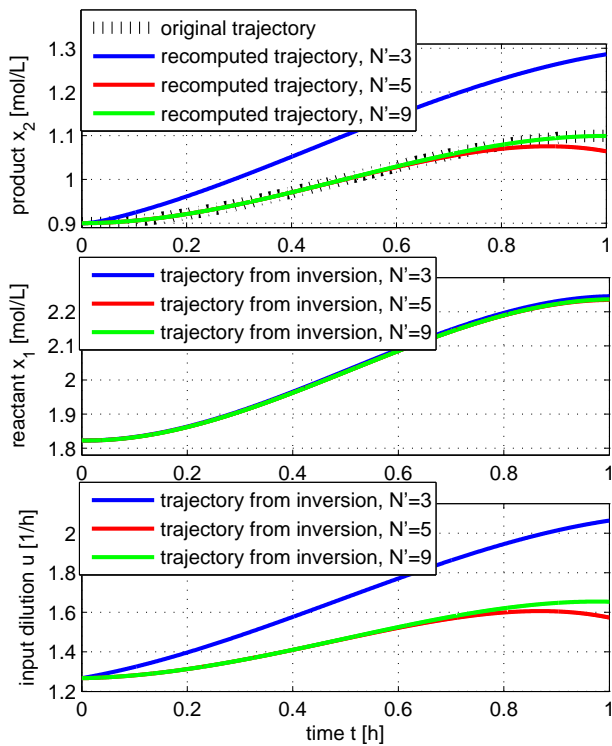}
  \caption{Van de Vusse type reactor: Results on inversion. Original product trajectory $x_2$, recomputed product trajectories from $u$ and (\ref{eq:VanVusse}) (Euler integration), reactant $x_1$ and input dilution $u$ computed by inversion, with $N'=3,5,9$.\label{fig:vanvusse}}
\end{figure}

A steady-state setpoint cange from $x_2=0.9$ to $x_2=1.1$ in $T=1$ h is considered. This fully determines the parameters $\alpha_i$, and the initial conditions define $\beta_0$. Furthermore, the system is in minimum-phase operation. The output reconstructed from the control input $u(t)$ by Euler integration is compared to the original output trajectory. Results for different orders $N'$ are displayed in Fig. \ref{fig:vanvusse}, namely $N'=3$, $N'=5$ and $N'=9$. In the results, $x_1(t)$ is always good as the impact of $u(t)$ is smaller than the impact of the output trajectory. The input, however, is converging slowly to the exact inversion input for an increasing order $N'$. For a quantitative comparison, the integrated trajectory offset 
\begin{align}
E &= \int_{t_0}^{t_f} \left| \ \tilde{y}(t) - y(t) \ \right| \ dt
\end{align}
is shown in table \ref{tbl:errors}. Although it seems that the solution is exact for $N'=9$, it is actually not, as there are higher-order terms in (50) as mentioned before. The found inverse models are usable only on a give time interval $t\in[t_0,t_f]$. Still, the results for $N'=9$ are exact enough for an application, as other error sources, such as parameter uncertainties, may have more impact as the error of the inverse model.

\begin{table}[h!]
\centering
\caption{Trajectory error for the inverse models\label{tbl:errors}}
\begin{tabular}{|l|c|c|c|}
\hline
degree $N'$ \;  & $3$ & $5$ & $9$  \\
\hline
error E     \;  &  $0.0964$ &   $0.0050$ & $0.00006$ \\
\hline
\end{tabular}
\end{table}

\section{CONCLUSIONS AND FUTURE WORKS}

\subsection{Conclusions}

In this paper, the problem of system inversion in the presence of a series-defined output trajectory is analyzed. Overall, the problem of system inversion is greatly simplified if the output is considered as a functional, which is a linear combination of basis functions with free coefficients. Known results on observability are used to establish ordinary differential equations. Subsequently, these are solved using series solutions and the method of \textit{undetermined coefficients} \cite{TP63}. An (algebraic) inverse model is found for an arbitrary given output, it consists of a set of equations for the system states and control input depending on time as well as on the (free) output trajectory coefficients.

An exact inverse model is found for linear systems. For general nonlinear systems, assuming that a system of given nonlinear equations can be solved, a good approximate inverse model valid on a finite time interval is found. 

The presented methods, compared to the existing differential-algebraic inversion schemes, have the advantage that the system is not required to be differentially flat (linearizable by feedback), and that the output is not redefined to the controller canonical form output \cite{HSR04}. Compared to the existing numerical schemes \cite{DCP96}, no differential equations are numerically integrated, which is an advantage if the system is of a high order or the trajectory horizon is high. A comparison to numerical methods is however not very useful, as the applying controller designs are different.

\subsection{Future Works}

For many applications, especially of the class of \textit{non-flat systems}, it is hard to obtain an inverse model or to design a feedforward controller \cite{HSR04}. If such a system is observable, it should be analyzed whether the method yields a suitable solution valid on a practicable time interval. Furthermore, in terms of robustness, it should be analyzed whether an approximate model is better than one with a redefined output.

\section{ACKNOWLEDGMENTS}

The authors thank the reviewers for their helpful and constructive comments.



\end{document}